\documentclass[reqno,12pt]{amsart}

\usepackage{fancyhdr,bbm}
\usepackage{datetime}
\settimeformat{ampmtime}
\usepackage[svgnames]{xcolor}
\usepackage{setspace,palatino}
\usepackage{pgf}
\usepackage{tikz}
\usepackage[scaled]{helvet} 
\usepackage{courier} 
\normalfont
\usepackage[T1]{fontenc}
\usepackage{booktabs}

\usetikzlibrary{patterns}
\usepackage{soul}
\usepackage{mathrsfs}
\usepackage{stmaryrd}
\usepackage[hidelinks=true]{hyperref}
\usepackage{natbib}
\usepackage{xfrac}
\usepackage[color=Orange]{todonotes}
\usepackage{layout}
\usepackage{microtype}
\usepackage{mathtools}
\usepackage{multirow}
\usepackage{placeins}
\usepackage{floatrow}
\usepackage{subfig}
\usepackage[multiple]{footmisc}

\newtheorem{lemma}{Lemma}

\newtheorem{theorem}{Theorem}

\usepackage{cleveref}

\crefname{figure}{figure}{figures}
\creflabelformat{figure}{#2#1#3}
\crefname{equation}{equation}{equations}
\creflabelformat{equation}{#2(#1)#3}
\crefname{lemma}{lemma}{lemmas}
\creflabelformat{lemma}{#2#1#3}
\crefname{theorem}{theorem}{theorems}
\creflabelformat{lemma}{#2#1#3}
\crefname{condition}{condition}{conditions}
\creflabelformat{condition}{#2#1#3}
\crefname{assumption}{assumption}{assumptions}
\creflabelformat{assumption}{#2#1#3}
\crefname{appendix}{appendix}{appendices}
\creflabelformat{appendix}{#2#1#3}
\crefname{enumi}{}{}
\creflabelformat{enumi}{(#2#1#3)}

\newtheorem{assumption}{Assumption}

\DeclareMathOperator{\argmax}{argmax}

\usepackage{amssymb}

\renewcommand{\Pr}{\mathbb{P}}

\usepackage{subfig}

\begin{document}

\thispagestyle{empty}

\title{Estimation of social--influence--dependent peer pressures  in a large network game}

\thanks{$^*$A previous version of this paper was circulated under the title ``Identification and Estimation of Hierarchy Effects in Social Interactions.'' We thank Anton Badev, Mark Hoekstra, Cheng Hsiao, Hae--shin Hwang, Jason Lindo, Xiaodong Liu, Essie Maasoumi, Steven Puller, Robin Sickles, Qi Li, Ken Wolpin, Mo Xiao, Keli Xu and participants at Econometric Society Summer Meeting 2014, Texas Econometrics Camp 2014, Emory, Monash, UT Austin and Texas A\&M for helpful feedbacks. We also thank Stephanie White for proofreading this paper. Supports from the ICPSR and Carolina Population Center of the National Longitudinal Study of Adolescent Health dataset  are gratefully acknowledged. }

\author{
\href{mailto:zhongjian.lin@emory.edu}{Zhongjian Lin$^\star$}}
\thanks{$^\star$Department of Economics, Emory University, Room 332, Rich Memorial Building, Atlanta, GA 30322,
\href{mailto:zhongjian.lin@emory.edu}{zhongjian.lin@emory.edu}}
\author{
\href{mailto:h.xu@austin.utexas.edu}{Haiqing Xu$^{\dag}$}}
\thanks{$^{\ddag}$Department of Economics, University of Texas at Austin, \href{mailto:h.xu@austin.utexas.edu}{h.xu@austin.utexas.edu}}

\date{\today}

\maketitle

\begin{abstract}
Research on peer effects in sociology has been focused for long on social influence power to investigate the social foundations for social interactions.  This paper extends \cite{xu2011social}'s large--network--based game model by allowing for social--influence--dependent peer effects.  In a large network, we use the Katz--Bonacich centrality to measure individuals' social influences. To solve the computational burden when the data come from the equilibrium of a large network, we extend \cite{aguirregabiria2007sequential}'s nested pseudo likelihood estimation (NPLE) approach to our large network game model. Using the Add Health dataset,  we investigate peer effects on conducting dangerous behaviors of high school students. Our results show that peer effects are statistically significant and positive. Moreover,  a student benefits more (statistically significant at the  5\% level) from her conformity, or equivalently, pays more for her disobedience, in terms of peer pressures, if friends have higher social-influence status. \\ 


\end{abstract}

\vspace{5ex}

\clearpage

\section{Introduction}

\label{sec:intro}

Game theoretic network models have been successful to study social interactions which has been traditional focuses of sociology. A leading example is network formation, e.g. \cite{jackson1996strategic,bala2000noncooperative} from the theory side. Another example considers social interactions in exogenously given large networks, e.g. \cite{blume2011linear,xu2011social}. In this paper, we extend \cite{xu2011social}'s large--network--based game model by allowing for social--influence--dependent peer effects. In particular, our research question is whether individuals of high social influence (measured by network centrality) impose more peer pressures to their followers than individuals of low social influence. 

Network positions are particularly important in studying all kinds of social interactions.  In sociology, researchers use network centrality, a concept introduced already in the late 1940's, to measure a social individual's  position, influence and prestige. In network--related policy analysis,  it is always disputable whether key players in a social network are more influential than ordinary indivdiuals simply due to their large number of followers and/or more central positions in the network,  i.e. the ``channel effects'' \citep[see e.g.][]{ibarra1993power, burt1995structural},  or because of their extraordinary influence ability on their  followers, i.e. the social influence effects. The answer to this question is crucial to evaluate e.g. targeting--and--remove key player policy \citep[][]{lee2012criminal}.

This paper builds upon \cite{xu2011social}'s large--network structural approach. In particular, we assume an individual's  payoff from her decision depends on her own covariates, as well as her direct friends' choices. As the fundamental principal in sociology, players benefits from choosing the same action of friends. There is an important substantive difference, however:    we allow peer's pressures/benefits for conformity to vary with friends' (relative) social influence/prestige, which is measured by  the Katz--Bonacich centrality \citep[see e.g.][]{katz1953new,bonacich1987power}. Such an extension is motivated from empirical applications on network--based policy analysis. Consider e.g. the targeting--and--remove key player policy. The constant peer pressure model will necessarily  underestimate key players' effects on their peers by diluting them with ordinary players' peer pressures, if there are (economically significant) influence effects on peer pressures. The inconsistent estimator of peer effects could further mislead the next stage counterfactual analysis of the policy. Moreover, in our empirical application, i.e., studying dangerous behaviors of high school students, the estimation results using  the Add Health dataset suggest that the constant peer pressure model should be too restrictive to provide consistent estimates. 

To our knowledge, only a handful of papers consider social influence status (measured by network centralities) in structural peer effects analysis.  In the spatial autoregressive model, \cite{calvo2009peer} develop a model that shows the Nash equilibrium outcome of each individual in the network is proportional to the individual's Katz--Bonacich centrality measure, which is assumed to capture all the direct and indirect influences of the network on a given individual.  An important empirical impaction of their approach is: the more central (in terms of Katz-Bonacich centrality) a person in a network, the higher level is her outcome (i.e. criminal activity). In contrast, we do not construct the network centrality measure from direct and indirect peer effects, but rather use the Katz--Bonacich centrality as an exogenous observable. Observations of such a measure directly obtain from the Add Health dataset.  Another important related paper is   \cite{liu2010gmm}, who use  the Katz--Bonacich centrality as an instrumental variable for peer effects in a linear social interaction model. In our structural approach, we assume that friends' Katz--Bonacich centralities affect peer pressures on a player and therefore affect her outcome directly.

To solve the computational burden for solving the equilibrium of a large network game, we apply \cite{aguirregabiria2007sequential}'s nested pseudo likelihood estimation (NPLE) approach to estimate our large network game model. It is a natural idea to extend  \cite{aguirregabiria2007sequential}' approach to large network games: Similar to dynamic games, because of the large dimensionality issue, it is costly to compute the equilibrium in a large network game using fixed point algorithms. The NPLE method starts with an arbitrary guess of the choice probabilities, e.g. the predicted choice probabilities from the standard Logit estimation without strategic interactions. Then we conduct another Logit estimation by using the predicted friends' choice probabilities as individual's expectation on friends' equilibrium behaviors. After that, we obtain an update of the predicted choice probabilities. We  repeat this updating procedure until it converges.  Therefore, NPLE is an iterative algorithm which consists of a sequence of Logit estimations. The contraction condition established in \cite{kasahara2012sequential} ensures the convergence of the algorithm. In a large social network game, the NPLE is attractive to practitioners due to simplicity of implementation and less time consuming.

Using the Add Health dataset,  we investigate peer effects on conducting dangerous behaviors of high school students. Our results show that peer effects are statistically significant and positive. Moreover, given friends chooses  ``not conducting dangerous behaviors'',  then a high school student should benefit more (statistically significant at the 5\% level) from her conformity, or equivalently, pays more for her disobedience, in terms of peer pressures, if friends have higher social-influence status. We also compare results from our model with  \cite{xu2011social}'s model and the standard Logit  model. In particular, the peer effects are insignificant in \cite{xu2011social}'s model. In the Logit model, coefficient estimate for friends' social influence status is negative and statistically significant at the 5\% level. Such a result suggests  a negative correlation between players' decisions and their friends social influence status. However, it is implausible to give it a meaningful economics interpretation.

The rest of the paper is organized as follows. Section \ref{sec:model} introduces our model and the definition of the Katz--Bonacich centrality. We also characterize the equilibrium  and establish its uniqueness. Section 3 establishes the identification of structural parameters and provides NPLE.  Asymptotic properties for NPLE are established and  finite sample performance is studied by Monte Carlo experiments.  Section 4  applies our estimation method to study peer effects of high school students on dangerous behaviors. Proofs of our results are collected in the Appendix.

%
%
%
%
%
\section{A Model of Social Interactions in Large Networks}
\label{sec:model}
We consider a discrete game played on an existing large social network. The network is viewed as a random graph with vertex connected with directed edges: In the graph, each individual $i\in\mathcal I\equiv \{1,\cdots,n\}$ is represented by a vertex, who is  connected to a group of best friends, represented by directed edges. Let $\ell_{ij}=1$ if individual $i$ nominates $j$ as a best friend, and $\ell_{ij}=0$ otherwise. Following convention, let $\ell_{ii}=0$ for all $i\in\mathcal I$. Moreover, we denote $F_i=\{j\in\mathcal I: \ell_{ij}=1\}$ as the group of  $i$'s best friends. By definition, best--friend relationship needs not be symmetric in our directed network; in other words, $\ell_{ij}\neq \ell_{ji}$ is allowed. Furthermore, we denote the network graph by an $n\times n$ matrix $\mathbb L$ where the $ij$--th entry is $\ell _{ij}$.

Using graph theory,  different metrics have been developed to quantify the influence of every node within a network \citep[see e.g.][]{borgatti2006graph}. In directed networks, for instance, \cite{knoke1983prominence} use the number of outgoing links and the number of incoming links to measure {\it influence} and {\it support}, respectively. Such kind of degree measures however are criticized for not taking into account those indirect connections to all the individuals in the network, but only immediate ones. Instead, we use Katz--Bonacich centrality measure as our social influence metric, as is suggested by e.g. \cite{bonacich1987power} in the sociology literature. Specifically, for $i=1,\cdots,n$, let 
\begin{equation}
\label{KB}
S_i=\sum_{k=1}^\infty\sum_{j=1}^n \lambda^{k}\times (\mathbb L^k)_{ji}.
\end{equation}where $\lambda\in(0,1)$ is the so--called attenuation factor. Note that $\sum_{j=1}^n (\mathbb L^k)_{ji}$ is the number of individuals who are $k$ steps away from $i$ in the network, where the network distance from $i$ to $j$ is defined as the smallest number of (directed) links that connects $i$ to $j$.   By definition, $S_i=0$ if $F_i=\emptyset$. 

\begin{figure}[h] 
   \centering
   \includegraphics[height=3.5in,width=5in]{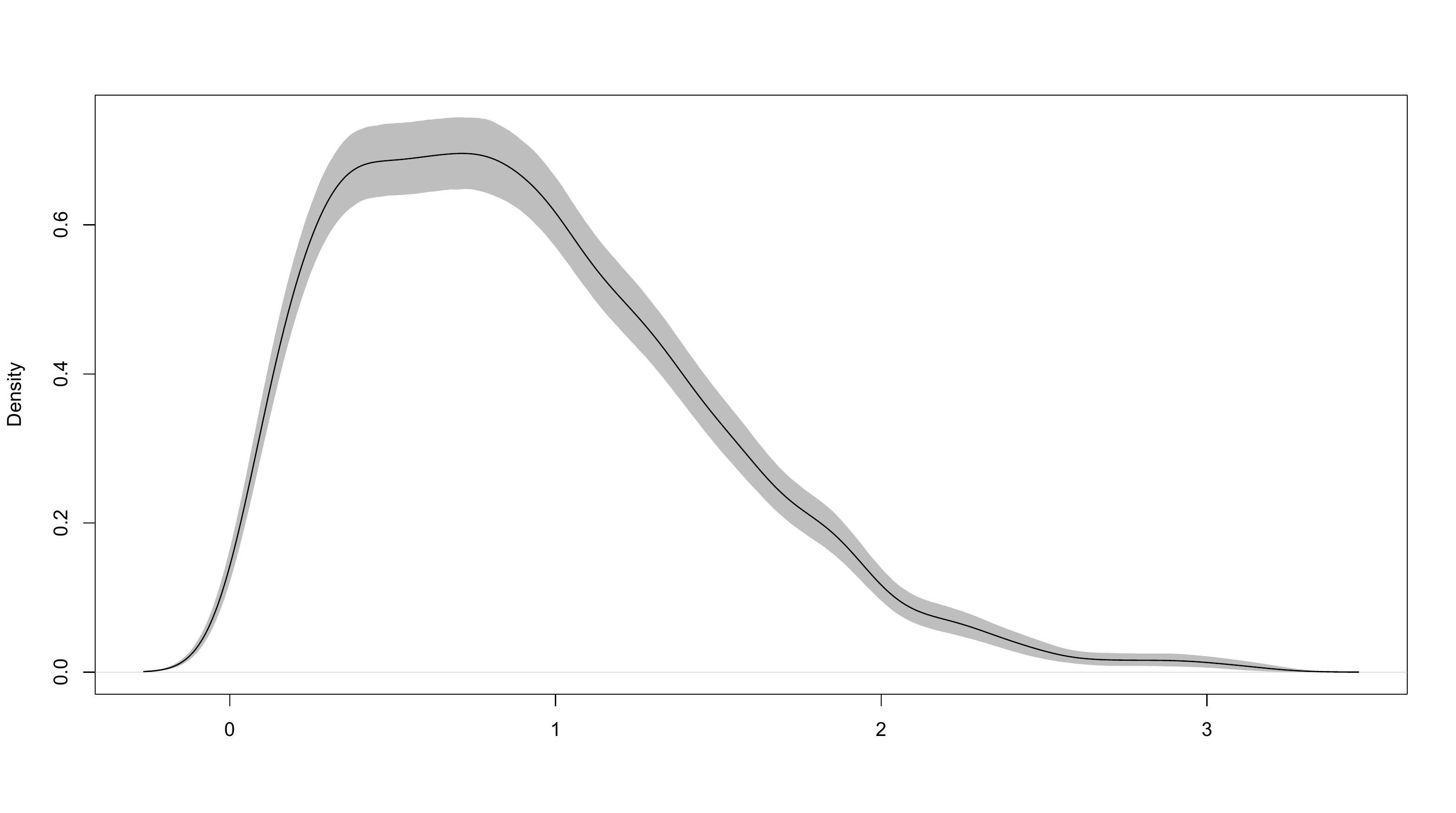} 
    \caption{Density function of Katz-Bonacich Centrality}
   \label{fig1}
\end{figure}

For our empirical application, the Add Health dataset contains such a measure with $\lambda=0.1$.   \Cref{fig1} provides a probability distribution of $S_i$ (conditional on $S_i>0$). In particular, the shadow area is the 95\% confidence interval. According to the picture, $S_i$ varies across individuals.  

In our network game, each individual simultaneously chooses $Y_i\in\{0,1\}$. In our empirical application, $Y_i=1$ refers to   student $i$  conducting dangerous behaviors in a recent period. Then the utility function of $i$ is given by
\begin{equation}
\label{eqn1}
U_{i}=\left\{\begin{array}{ll}X'_i\beta_1+\frac{1}{Q_i}\sum_{j\in F_i} \alpha_1(S_j-S_i)\times \mathbbm 1(Y_j=1) -\epsilon_{1i}, & \text{if } Y_i=1; \\X'_i\beta_0+\frac{1}{Q_i}\sum_{j\in F_i} \alpha_0(S_j-S_i)\times \mathbbm 1(Y_j=0) -\epsilon_{0i}, & \text{if }Y_i=0,\end{array}\right.
\end{equation} where $X_i\in \mathbb R^d$ includes a constant and a vector of individual $i$'s demographic characteristics, $\epsilon_{0i},\epsilon_{1i}\in\mathbb R$ are  unobserved action--dependent utility shocks, and $Q_i\equiv \sum_{j=1}^n \mathbb L_{ji}=\sum_{j=1}^n  \mathbbm{1}(j\in F_i)$ denotes the total number of friends. For expositional simplicity, we assume $Q_i\geq 1$ in  \eqref{eqn1}.\footnote{The case of $Q_i=0$ can be  accommodated simply by letting $U_{i}=X_i'\beta_d$ if $Y_i=d$.} Moreover, $\beta_1,\beta_2\in\mathbb R^d$ are  payoff coefficients,  $\alpha_0(\cdot)$ and $\alpha_1(\cdot)$ are unknown structural functions. In particular, $\alpha_1$ measures peer pressures on player $i$ from her friend $j$ choosing the same action of dangerous behaviors, i.e. $Y_i=Y_j=1$.   A similar interpretation applies to $\alpha_0$. Furthermore, $\alpha^\dag(S_j-S_i)\equiv \alpha_1(S_j-S_i)+\alpha_0(S_j-S_i)$ measures the total peer effects that lead to  conformity among friends in a social network.  It is worth pointing out that the payoff of action 0 is not ``normalized'' to  zero. In our influence--dependent social interaction model,  zero--payoff for action 0 is not an innocuous normalization for reasons to be discussed later. See \cite{buchholz2016semiparametric} for a similar but more detailed argument.

In the above payoff function,  a key feature is that  $\alpha_1(S_j-S_i)$ and $\alpha_0(S_j-S_i)$ depend on  friend $j$'s  relative social influence $S_j-S_i$. Such a specification on peer effects is related to social influence models used in sociology, e.g. \cite{friedkin1990social}. In our empirical application, we investigate the question, if an individual's friends choose (not) to conduct dangerous behaviors, whether the amount of peer pressures from friends increases with friends' social influence.

Next, we  specify the information structure: Let $X_i$ and $\mathbb L$ be publicly observed state variables, and $(\epsilon_{0i},\epsilon_{1i})$ be player $i$'s private information. Note that because $F_i$ and $S_i$ are derived from $\mathbb L$, therefore they are also publicly observed  state variables. Let $\mathbb W=\{(X'_1,\cdots,X'_n)'; \mathbb L\}$ be all the public information in the game. According to the  {\it Bayesian Nash Equilibrium} (BNE) solution concept,  the best response function is given by 
\[
R_i(\mathbb W, \epsilon_i)
= \mathbbm 1\left\{ X_i'(\beta_1-\beta_0)-\sum_{j\in F_i}\frac{\alpha_0(S_j- S_i)}{Q_i}+\sum_{j\in F_i}\frac{\alpha^\dag(S_j- S_i)\Pr(Y_j=1|\mathbb W)}{Q_i} -\epsilon^*_{i}\geq 0\right\},
\]where  $\epsilon^*=\epsilon_1-\epsilon_0$.  In equilibrium, players decisions can be written by 
\[
Y_i=R_i(\mathbb W,\epsilon_i), \ \ \forall i\leq n.
\]

\subsection{Equilibrium Characterization}
To characterize the equilibrium, we first make an assumption on the distribution of $\epsilon_i$. 
\begin{assumption}
\label{ass1}
The error terms $\{(\epsilon_{0i},\epsilon_{1i}): i\leq n\}$ are distributed i.i.d. across both actions and players.
Furthermore, the error term has an extreme value distribution with density
\[
f(t)=\exp(-t)\exp(-\exp(-t)). 
\]
\end{assumption}
\noindent 
\Cref{ass1} is fairly standard in discrete choice model literature, e.g. \cite{bajari2012estimating}. As a matter of fact, \Cref{ass1} provides a closed--form expression for players' best responses in terms of choice probabilities.

Let $\sigma^*_i(\mathbb W)=\Pr(Y_i=1|\mathbb W)$ be the equilibrium choice probability of choosing action one. Let further $\bar \alpha_{0i}(\cdot)=\alpha_0(\cdot)/Q_i$ and $\bar\alpha^\dag_i(\cdot)=\alpha^\dag(\cdot)/Q_i$. Under \Cref{ass1}, the best response function can be written in terms of equilibrium choice probabilities, i.e., for $i=1,\cdots,n$,
\begin{equation}
\label{choice_prob}
\sigma^*_i(\mathbb W)
= \frac{ \exp\left\{X_i'(\beta_1-\beta_0)-\sum_{j\in F_i} \bar \alpha_{0i}(S_j-S_i)+\sum_{j\in F_i}\bar\alpha^\dag_i(S_j- S_i) \sigma^*_j(\mathbb W)\right\}}{ 1+\exp\left\{X_i'(\beta_1-\beta_0)-\sum_{j\in F_i} \bar \alpha_{0i}(S_j-S_i)+\sum_{j\in F_i} \bar\alpha_i(S_j- S_i) \sigma^*_j(\mathbb W)\right\}}.
\end{equation}

To ensure the equation system \eqref{choice_prob} admits a unique solution, we next introduce an assumption on the strength of peer effects. Let $\mathcal S_0$ be the support of $S_j-S_i$ where  $j\in F_i$.
\begin{assumption}
\label{ass2}
Let $  \underset{s\in \mathcal S_0}\sup \ \big|\alpha_0(s)+\alpha_1(s)\big|< 4$.
\end{assumption}
\noindent
Under \Cref{ass2}, the dependence of the equilibrium choices satisfies the mixing conditions, which serve as a key to dependent data analysis. Similar assumptions  for equilibrium uniqueness in Bayesian games can also be found in e.g. \cite{brock2001discrete,horst2006equilibria} and \cite{xu2011social}. 

\begin{theorem}
\label{thm1}
Under \cref{ass1,ass2}, there exists a unique pure strategy BNE for any $n$.
\end{theorem}
\noindent
\Cref{thm1} is important for statistical inference on large network games. When there are multiple equilibria, an obvious obstacle for statistical inference is the incompleteness of the econometrics model. For more discussions on issues of multiple equilibria, see e.g. \cite{tamer2010partial,tamer2003incomplete} and \cite{de2013econometric}.

\section{Identification and Estimation}
For tractability, we linearize $\alpha(\cdot)$ for our empirical analysis: Let $\alpha_0(s)=\phi_0+\phi_1\times s$ and $\alpha_1(s)=\psi_0+\psi_1\times s$.  By definition, $\alpha^\dag(s)=\alpha_0(s)+\alpha_1(s)=\phi_0+\psi_0+ (\phi_1+\psi_1) \times s$. By the equilibrium condition \eqref{choice_prob},  $\beta_1$ and $\beta_0$ cannot be separately identified in the structural model, since only their difference $\beta_1-\beta_0$ matters for the equilibrium.  Therefore, we set $\beta_0=0$ as a normalization. 

\subsection{Identification}
 First, note that $\sigma^*_i(\mathbb W)$  obtains directly from the distribution of observables. Following the definition of identification \citep[see e.g.][]{hurwicz1950generalization},  we treat $\sigma^*_i(\mathbb W)$  as a known object.\footnote{See \cite{xu2011social} for a detailed discussion on the definition of identification in a single large network game model.} Let $T_i=\ln\sigma^*_i(\mathbb W)-\ln [1-\sigma^*_i(\mathbb W)]$. It follows by \eqref{choice_prob} that 
\[
T_i=X_i'\beta_1-\phi_0-\phi_1\frac{ \sum_{j\in F_j}\bar S_{ji}}{Q_i}+(\phi_0+\psi_0)\frac{\sum_{j\in F_i} \sigma^*_j(\mathbb W)}{Q_i}
+(\phi_1+\psi_1)\frac{\sum_{j\in F_i} \bar S_{ji}\sigma^*_j(\mathbb W)}{Q_i},
\]
where $\bar S_{ji}=S_j-S_i$.  Note that $\phi_0$ cannot be separately identified  from the constant term of $X'\beta_1$. Therefore, let $\phi_0=0$ as a normalization. It follows that
\[
T_i=X_i'\beta_1-\phi_1\frac{\sum_{j\in F_i} \bar S_{ji}[1-\sigma^*_j(\mathbb W)]}{Q_i}
+\psi_0\frac{\sum_{j\in F_i} \sigma^*_j(\mathbb W)}{Q_i}+\psi_1\frac{\sum_{j\in F_i} \bar S_{ji}\sigma^*_j(\mathbb W)}{Q_i},
\] which takes a linear expression of structural parameters. 

Let  $\theta\equiv (\beta_1,\phi_1,\psi_0,\psi_1)\in\Theta \subseteq \mathbb R^{d}\times \mathbb R^3_+$, where $ \Theta$ is  the parameter space. The positiveness of $\phi_1,\psi_0$ and $\psi_1$ reflects the fundamental principal in sociology that friends benefit from conformity.  Let $\theta_0$ and $\theta$ be the true parameter for the data generating process and a generic value in $\Theta$, respectively.  Moreover, we denote 
\[
Z_i\equiv \left(X_i', \frac{\sum_{j\in F_i}\bar S_{ji}[\sigma^*_j(\mathbb W)-1]}{Q_i},
  \frac{\sum_{j\in F_i}\sigma^*_j(\mathbb W)}{Q_i}, \frac{\sum_{j\in F_i}\bar S_{ji}\sigma^*_j(\mathbb W)}{Q_i}\right)'\in\mathbb R^{d+3}.
\]

\begin{assumption}
\label{ass3}
$\mathbb E (Z_iZ_i')$ have full rank, i.e., $\text{Rank}\left(\mathbb E (Z_iZ_i')\right)=d+3$. 
\end{assumption}
\noindent
\Cref{ass3} is  a high--level rank condition that  requires  no perfect collinearity of $Z_i$. Such a full rank condition can hold if (i) $X_i$ has no perfect collinearity; (ii) conditional on $(X_i, S_i,Q_i,\{ S_{j}: j\in F_i\})$, $\{\sigma^*_j(\mathbb W): j\in F_i\}$ has no perfect collinearity; (iii) For every $j\in F_i$, conditional on $(X_i,  F_i/\{j\})$, we have $0<\mathbb P(j\in F_i)<1$.  In particular, the last condition requires variations in $Q_i$ given $X_i$.  Note that \Cref{ass3} is testable given $Z_i$ can be nonparametrically estimated \citep[see][]{xu2011social}.
\begin{lemma}
\label{lemma1}
Suppose \Cref{ass1,ass2,ass3} hold.  Then $\theta_0$ is identified by
\begin{equation}
\label{eqn5}
\Big[\mathbb E\big(Z_iZ_i'\big)\Big]^{-1}\mathbb E\big(Z_iT_i\big).
\end{equation}
\end{lemma}
\noindent
The proof directly follows our discussions above, and hence is omitted. 

It is worth pointing out that if friends' relative social influences have a strictly positive effect on peer pressures, i.e., $\phi_1>0$ and/or $\psi_1>0$, then ignoring such 
an effect will necessarily induce omitted variable bias to the estimation of peer effects. To see this, suppose equilibrium beliefs $\{\sigma_j^*(\mathbb W): j=1,\cdots,n\}$ are observed in the data. Then a Logit estimation without including $\frac{\sum_{j\in F_i}\bar S_{ji}[\sigma^*_j(\mathbb W)-1]}{Q_i}$ and $\frac{\sum_{j\in F_i}\bar S_{ji}\sigma^*_j(\mathbb W)}{Q_i}$ as regressors would be inconsistent due to their correlation with $\frac{\sum_{j\in F_i}\sigma^*_j(\mathbb W)}{Q_i}$, the regressor for the constant peer effect coefficient $\psi_0$.

\subsection{Estimation}
\label{sec:est}
Our estimation follows \cite{aguirregabiria2007sequential}'s NPLE approach. Similarly to their dynamic setting, the difficulties in large network games arise from the computational burden of solving the equilibrium. Using an iterative algorithm, the NPLE significantly reduces the computational burden, albeit it is less efficient than the maximum (pseudo) likelihood estimation approach \citep[see][]{xu2011social}. More importantly, the proposed approach is  essentially a sequence of  Logit estimations., which is easy to implement.

Consider a random sample $\{(Y_i, X_i,F_i):i=1,\cdots,n\}$ from a single large social network. It is worth pointing out that our approach can be easily extended to applications where observations come from a small number of networks but each network has a large size. In both cases, our asymptotic analysis relies on the number of players going to infinity.    

Under this parametric specification, we are particularly interested in testing $\mathbb H_0: \phi_1=\psi_1=0$  versus $\mathbb H_1: \phi_1\neq 0\ \text{ or }\ \psi_1\neq 0$. In such a significance test, rejection of the null hypotheses provides evidence for causal effects from friends' social influence on peer pressures. 

\subsection{NPLE algorithm}
Let $\Sigma^*(\mathbb W)=\big(\sigma_1^*(\mathbb W), \cdots, \sigma_n^*(\mathbb W)\big)'$ and $\Sigma=(\sigma_1,\cdots,\sigma_n)'\in [0,1]^n$ be the equilibrium choice probability profile and a generic probability profile, respectively.  For arbitrary $\Sigma\in[0,1]^n$, let
\[
 Z_i(\Sigma)\equiv \left(X_i', \frac{\sum_{j\in F_i}\bar S_{ji}(\sigma_j-1)}{Q_i},
  \frac{\sum_{j\in F_i}\sigma_j}{Q_i}, \frac{\sum_{j\in F_i}\bar S_{ji}\sigma_j}{Q_i}\right)'
\]
and
 \[
\Gamma_i(\Sigma,\theta; \mathbb W)= \frac{ \exp[Z_i'(\Sigma)\theta]}{ 1+\exp[Z'_i(\Sigma)\theta]}.
\] Moreover, we denote $\Sigma(\theta; \mathbb W)$ as the solution to the equation system: 
\[
\Gamma_i(\Sigma, \theta;\mathbb W)=\sigma_i, \ \ \forall \ i\leq n.
\] By definition, $\Sigma^*(\mathbb W)=\Sigma(\theta_0;\mathbb W)$. Furthermore,   let
\begin{align*}
&\hat L_n(\theta, \Sigma)=\frac{1}{n}\sum_{i=1}^n\left\{Y_i\ln \Gamma_i (\Sigma, \theta;\mathbb W)+(1-Y_i)\ln \left[1-\Gamma_i(\Sigma,\theta; \mathbb W)\right]\right\};\\
&L(\theta, \Sigma)=\lim_{n\rightarrow \infty}\frac{1}{n}\sum_{i=1}^n\mathbb E \left\{Y_i\ln \Gamma_i (\Sigma, \theta;\mathbb W)+(1-Y_i)\ln \left[1-\Gamma_i(\Sigma,\theta; \mathbb W)\right]\right\}.
\end{align*}
Note that $L(\theta,\Sigma)$ is defined as the limiting log-likelihood function as the network size goes to infinity. 

Given the above notation, we are ready to describe our estimation procedure: First, we start with an arbitrary initial value $\Sigma^{[0]}\in[0,1]^n$, w.l.o.g., let  $\Sigma^{[0]}=(0,\cdots,0)\in [0,1]^n$. Next, we iterate the following two steps:

\noindent
Step 1. Given $ \Sigma^{[j-1]}$, let
\[
\hat \theta^{[j]}= \argmax_{\theta\in\Theta} \ \hat L_n(\theta, \Sigma^{[j-1]}).
\]

\noindent
Step 2. Given $\hat \theta^{[j]}$, let 
\[
 \Sigma^{[j]}=\Gamma(\Sigma^{[{j-1}]}, \hat \theta^{[j]}; \mathbb W),
\]where $ \Gamma(\Sigma,\theta;\mathbb W)=\big( \Gamma_1(\Sigma,\theta;\mathbb W),\cdots, \Gamma_n(\Sigma,\theta;\mathbb W)\big)'$.
This procedure stops at the $K$--th iteration when $\|\hat\theta^{[K]}-\hat\theta^{[K-1]}\|$ is less than a predetermined tolerance, e.g., $10^{-6}$. Then, we define our estimator by $\hat \theta_{NPLE}=\hat \theta^{[K]}$. The convergence of the NPLE algorithm is ensured by the local contraction condition established in \cite{kasahara2012sequential}. 

By definition, the above NPLE is essentially a fixed point solution to maximize the log-likelihood function, which  can be equivalently defined by 
\begin{align*}
&\hat \theta_{NPLE}= \argmax_{\theta\in\Theta} \hat L_n(\theta,\Sigma),\\
& s.t.\ \ \    \Sigma= \Gamma(\Sigma,\theta;\mathbb W).
\end{align*} See e.g. \cite{aguirregabiria2007sequential} for a more detailed discussion.

\subsection{Asymptotic analysis}
We make further assumptions to establish asymptotic properties of $\hat \theta_{NPLE}$.
\begin{assumption}
\label{ass5}
The underlying parameter $\theta_0$ uniquely solves the following equation:
\[
 \theta=\argmax_{c\in\Theta} L(c , \Sigma( \theta;\mathbb W)).
\]
\end{assumption}
\noindent
\Cref{ass5} is essentially an identification assumption for the pseudo log--likelihood function $L(c , \Sigma( \theta;\mathbb W))$. It is straightforward that $\theta_0$ solves the moment equation: $\theta_0$ maximizes $L(\cdot,\Sigma^*(\mathbb W))$ by the standard argument for MLE and note that $\Sigma^*(W)=\Sigma(\theta_0;\mathbb W)$. The true parameter $\theta_0$ being the unique solution can be ensured if the function $\argmax_{c\in\Theta} L(c , \Sigma( \theta;\mathbb W))$ is a contraction mapping from $\Theta$ to $\Theta$.\footnote{It is worth pointing out that in the proof of \Cref{thm1} players' best responses are shown to be a contraction mapping in the space of strategy profiles, but not in the parameter space. }  Such a condition highlights the identification power of the ``true'' log--likelihood: either $L(\cdot , \Sigma(\cdot;\mathbb W))$ derived from the game structure, or $L(\cdot , \Sigma^*(\mathbb W))$ that depends on unobserved equilibrium beliefs $\Sigma^*(\mathbb W)$. When the moment equation in \Cref{ass5} is sufficient for (global) identification of $\theta_0$, then the MLE approach can be simplified by the NPLE algorithm. 

\Cref{ass5} can be verified by the data. If it fails, as  \cite{aguirregabiria2007sequential} suggest, the NPLE algorithm  should be modified by selecting the fixed point that maximizes the value of the pseudo likelihood.

\begin{assumption}
\label{ass6}
$\mathscr S_X$ is bounded and $\Theta$ is compact. 
\end{assumption}
\noindent
\Cref{ass6} ensures that choice probabilities derived from the model are uniformly bounded away from zero, which implies that $\hat L_n(\cdot,\Sigma^{[j]})$ is also uniformly bounded for all $j$.

\begin{assumption}
\label{ass7}
Let $\max_i\in\{1,\cdots,n\} \sum_{j=1}^n\ell_{ij}\leq M$ for some constant $M\in\mathbb N^+$.
\end{assumption}
\noindent
\Cref{ass8} is needed to limit the dependence among all the observations. In our Add Health dataset,  $M=10$. 

For any $h\in\mathbb N$ and $i\in\mathcal I$, let $N_{(i,h)}=\{j\in\mathcal I: (\mathbb L^k)_{ji}=1\ \text{ for some } k\leq h\}$. Moreover, let $  \mathbb L^{(i,h)}$ be a $\#N_{(i,h)}\times \#N_{(i,h)}$ submatrix of $\mathbb L$ which describes the graph for the subnetwork among  $N_{(i,h)}$.

\begin{assumption}
\label{ass8}
Fix arbitrary $h\in\mathbb N$.  The probability distribution of $\mathbb L^{(i,h)}$ converges to a limiting distribution  as $n\rightarrow \infty$ for all $i$; and $\mathbb L^{(i,h)}$ is independent of  $\mathbb L^{(j,h)}$ if $ N_{(i,h)}\cap N_{(j,h)}=\emptyset$. Moreover, the payoff covariates $X_i$ are i.i.d. across players given the exogenous random network. 
\end{assumption}
\noindent
In the large network asymptotics, \Cref{ass8} is also made in \cite{xu2011social} for the consistency of an MLE--type estimator. In particular, this condition requires that the distribution of subgraphs should converge to a limit as the network size goes to infinity, and two non--overlapping subgraphs have independent connecting structures. 

\begin{theorem}
\label{thm2}
Suppose \Cref{ass1,ass2,ass3,ass5,ass6,ass7,ass8} hold. In particular, \Cref{ass2} holds for all $\theta\in\Theta$. Then
\[
\hat \theta_{NPLE}\overset{p}{\rightarrow} \theta_0.
\]
\end{theorem}
\noindent
In \Cref{thm2}, we need restrict the parameter space for $\alpha(\cdot)$ such that  \Cref{ass2} holds for all $\theta\in \Theta$. Similar to the stationary restriction in the autoregressive model, such a condition imposes restrictions on $(\phi_1,\psi_0,\psi_1)$ that depend on the support of $S_j-S_i$. 

Following \cite{aguirregabiria2007sequential}, we now derive the limiting distribution of $\hat \theta_{NPLE}$.   Let $A_n=\mathbb E \big[Z_iZ_i'\sigma_i^*(\mathbb W)(1-\sigma_i^*(\mathbb W))+\frac{1}{Q_i} \sum_{j\in F_i}Z_i Z_j'\sigma^*_j(\mathbb W)(1-\sigma^*_j(\mathbb W))(\psi_0+(\phi_1+\psi_1) \bar S_{ji})\big]$ and $B_n= \mathbb E \big[Z_iZ_i'(Y_i-\sigma^*_i(\mathbb W))^2\big]$. Note that $A_n$ and $B_n$ depend on index $n$ through $\mathbb W$.

\begin{assumption}
\label{ass9}
$\theta_0$ belongs to the interior of $\Theta$. 
\end{assumption}
\begin{assumption}
\label{ass10}
There exist non--singular $(R+3)\times (R+3)$ matrix $A$ and $B$ such that $A_n\rightarrow A$ and $B_n\rightarrow B$.
\end{assumption}
\noindent
\Cref{ass9} is standard in the asymptotics theory.  \Cref{ass10} is a high level condition which requires (i) $A_n$ and $B_n$ converge to some non--singular limiting matrices respectively as the network size goes to infinity. Such a condition could be derived by specifying a network growing mechanism. Moreover,  the non--degeneracy of $A$ and $B$ requires that all the determinants of $A_n$ and $B_n$ should be outside of an open ball of zero for all $n$, which is essentially a rank condition. 
\begin{theorem}\label{thm3}
Suppose all the conditions in \Cref{thm2} and \Cref{ass9,ass10} hold. Then we have
\[
\sqrt n(\hat\theta_{NPLE}-\theta_0)\overset{d}{\rightarrow} N\big(0,\Omega_0\big),
\]
where $\Omega_0=A^{-1} B A^{-1}$.
\end{theorem}

\subsection{Monte Carlo Experiments}
\label{sec:monte}
In this section, we investigate the finite sample performance of our estimator by using Monte Carlo experiments. First, we simulate a large network: The number of friends $Q_i$ of each player is drawn uniformly from $\{0,1,2,\cdots, 10\}$. Given $Q_i$, player $i$ randomly nominates her friends among all individuals for generating $F_i$. Moreover, we set $X_i=(1, W'_i,S_i)'\in\mathbb R^5$ where $W_i$ consists of three independent elements: The first element is uniformly distributed on $[-\sqrt 3,\sqrt 3]$, the second is a standard normal random variable, and the last conforms to a transformed Bernoulli distribution $2\times B(0.5)-1\in\{-1,1\}$. In this setting, every element of $W_i$ has mean zero and variance one. Following the Add Health dataset, we calculate the Katz-Bonacich centrality measure $S_i$ by \eqref{KB} with $\lambda=0.1$. In this specification,  $\beta=(\beta_0,\beta_1,\beta_2,\beta_3,\beta_4)'\in\mathbb R^5$ and $(\phi_1,\psi_0,\psi_1)\in\mathbb R^3_+$. 

Furthermore, we choose sample size $n = 400, 800$, and $1600$. All results are drawn from 1000 replications. We set $\beta=(-1,1,-1,1,-1)'$, $(\phi_1,\psi_0,\psi_1)=(1,1,1)'$, $(1,1,0)'$, $(0,1,0)'$, and $(1,2,0)'$, respectively.\footnote{We also consider other values of $\beta$ and $(\phi_1,\psi_0,\psi_1)$. The results are qualitatively similar.}  \Cref{table2,table3} report the finite sample performance of $\hat\theta_{NPLE}$, including average bias, standard deviation (in parentheses) and mean square error. In all of our experiments, the estimator behaves well. In particular, the mean squared error  decreases at the rate $n$ as the sample size increases.
\begin{center}
\begin{table}
\small
\caption{Simulation Results: Average Bias and Standard Deviation}
\label{table2}
\begin{tabular}{cc|cccccccc}
\hline
$(\phi_1,\psi_0,\psi_1)$     &$n$   &$\beta_0$&$\beta_1$&$\beta_2$&$\beta_3$&$\beta_4$&$\phi_1$ & $\psi_0$&$\psi_1$\\\hline
$(1,1,1)$&400&-0.012&0.040&-0.028&0.045&-0.044&0.036&-0.010&-0.097\\
&&(0.698)&(0.175)&(0.183)&(0.175)&(0.587)&(0.742)&(1.163)&(1.960)\\
&800&0.010&0.019&-0.016&0.019&-0.041&0.005&-0.015&-0.119\\
&&(0.426)&(0.124)&(0.120)&(0.117)&(0.381)&(0.479)&(0.755)&(1.322)\\
&1600&-0.014&0.011&-0.010&0.010&-0.004&-0.010&-0.016&0.024\\
&&(0.335)&(0.081)&(0.088)&(0.085)&(0.288)&(0.359)&(0.549)&(0.900)\\\hline
 $(1,1,0)$&400&-0.016&0.044&-0.030&0.049&-0.045&0.046&0.004&-0.080\\
&&(0.704)&(0.176)&(0.185)&(0.178)&(0.594)&(0.749)&(1.159)&(1.955)\\
&800&0.008&0.019&-0.016&0.020&-0.039&0.004&-0.022&-0.127\\
&&(0.427)&(0.123)&(0.120)&(0.119)&(0.384)&(0.478)&(0.746)&(1.330)\\
&1600&-0.015&0.011&-0.011&0.010&-0.001&-0.010&-0.023&0.011\\
&&(0.333)&(0.081)&(0.087)&(0.085)&(0.290)&(0.363)&(0.551)&(0.914) \\\hline
$(0,1,0)$&400&-0.014&0.040&-0.030&0.042&-0.035&0.016&-0.030&-0.030\\
&&(0.704)&(0.175)&(0.174)&(0.171)&(0.590)&(0.708)&(1.102)&(1.943)\\
&800&0.007&0.018&-0.018&0.019&-0.038&-0.006&-0.020&-0.119\\
&&(0.432)&(0.123)&(0.121)&(0.114)&(0.387)&(0.454)&(0.718)&(1.308)\\
&1600&-0.011&0.008&-0.008&0.010&-0.004&-0.009&-0.026&0.027\\
&&(0.336)&(0.078)&(0.084)&(0.082)&(0.290)&(0.348)&(0.521)&(0.900)\\\hline
$(1,2,0)$&400&-0.012&0.038&-0.030&0.041&-0.041&0.032&0.016&-0.061\\
&&(0.659)&(0.169)&(0.176)&(0.167)&(0.546)&(0.746)&(1.019)&(1.695)\\
&800&0.005&0.015&-0.016&0.016&-0.031&-0.002&-0.003&-0.101\\
&&(0.416)&(0.120)&(0.118)&(0.113)&(0.373)&(0.486)&(0.663)&(1.178)\\
&1600&-0.015&0.009&-0.009&0.010&-0.002&-0.013&0.000&-0.010\\
&&(0.315)&(0.079)&(0.085)&(0.082)&(0.272)&(0.363)&(0.486)&(0.788)\\\hline
\end{tabular}
\end{table}
\end{center}

\begin{center}
\begin{table}
\small
\caption{Simulation Results: Mean Square Error}
\label{table3}
\begin{tabular}{cc|cccccccccc}
\hline
$(\phi_1,\psi_0,\psi_1)$     &$n$   &$\beta_0$&$\beta_1$&$\beta_2$&$\beta_3$&$\beta_4$&$\phi_1$ & $\psi_0$&$\psi_1$\\\hline
$(1,1,1)$&400&0.487&0.032&0.034&0.033&0.346&0.552&1.352&3.848\\
&800&0.181&0.016&0.015&0.014&0.146&0.229&0.570&1.761\\
&1600&0.112&0.007&0.008&0.007&0.083&0.129&0.301&0.809\\\hline
$(1,1,0)$&400&0.496&0.033&0.035&0.034&0.355&0.563&1.341&3.824\\
&800&0.183&0.015&0.015&0.015&0.149&0.228&0.556&1.784\\
&1600&0.111&0.007&0.008&0.007&0.084&0.132&0.304&0.834\\
\hline
$(0,1,0)$&400&0.495&0.032&0.031&0.031&0.349&0.502&1.215&3.774\\
&800&0.186&0.016&0.015&0.013&0.151&0.206&0.516&1.724\\
&1600&0.113&0.006&0.007&0.007&0.084&0.121&0.272&0.809\\\hline
$(1,2,0)$&400&0.434&0.030&0.032&0.030&0.299&0.557&1.039&2.873\\
&800&0.173&0.015&0.014&0.013&0.140&0.236&0.439&1.396\\
&1600&0.100&0.006&0.007&0.007&0.074&0.132&0.236&0.620\\\hline
\end{tabular}
\end{table}
\end{center}

\section{Empirical application: Dangerous Behaviors of high school students}
\label{sec:risky}
In this section we apply our method to study  peer effects on the dangerous behaviors of high school students. Adolescent risky behaviors have been studied in terms of peer effects \cite[see e.g. ][]{nakajima2007measuring,gaviria2001school}. To the best of our knowledge, however, there is no structural analysis on social influence dependent peer effects in the current literature. In particular, the research question we ask is how  students of high social influence status, who  typically are less likely to conduct dangerous behaviors, affect their peers through the network. There is no doubt that they should affect more people given their high centrality in the network, but do they impose more peer pressures than ordinary peers do to their followers (i.e. direct friends)? In this paper, we use the self-report questionnaires from the \textit{National Longitudinal Study of Adolescent Health (Add Health)} dataset to study this empirical question.
\subsection{Add Health Dataset}
\label{ssec:data}
The  \textit{Add Health} is a longitudinal study of a nationally representative sample of adolescents in grades 7-12 in the United States during the 1994-95 school year. \textit{Add Health} combines longitudinal survey data on respondents' social and economic features with contextual data on the family, friendships and peer groups. In the dataset, each student has nominations of at most five male friends and at most five female friends, which allows us to construct a social network among observations. From the {\it Wave I} survey, we have 85,627 students from more than 100 representative schools in all regions of the united states. In this study, we pick a pair of sister schools, i.e.  No. 62 and No. 162,  with a significant proportion of inter-school friend nominations. Our sample contains 2,460 students. \Cref{table5} provides summary  statistics of the observables. 

\begin{table}
\small
\caption{Summary of Statistics of Key Variables from the Data}
\label{table5}
\begin{center}
\begin{tabular}{lcccc}
\hline
Variable& Min & Max & Mean& Std. Deviation \\\hline
Risky Behaviors   &0&1&0.44&    0.50\\
Age                      &10&19&15.18 &1.64\\
Female                &0&1&0.52&    0.54\\
KB Centrality       &0&3.14&0.82&0.60\\
Ave. Friends' KB Centrality &0&2.68&0.84&0.49\\
Number of friends      &0 &10 &4.86&2.86\\\hline 
\end{tabular}
\end{center}
\end{table}
Observed demographic characteristics include age and gender, as well as the Katz-Bonacich (KB) centrality measure. The average friends' KB centrality measure is constructed from the friends nominations. The dependent variable is constructed by using   the self-report questionnaires in the Add Health dataset. Specifically, the survey question is ``During the past twelve months, how often did you do something dangerous because you were dared to?''

\subsection{Empirical Results}

\Cref{table6} reports our estimate results. Clearly, male students are more ``dared to'' do dangerous things than female students, and students of higher social influence status are less likely to conduct dangerous behaviors. The effects of age however are not significant. Moreover, we find significant peer pressures on a player of choosing the same action, when his friends choose dangerous behaviors. However, such peer effects are insignificantly affected by friends' relative social influence status. On the other hand, when his friends choose to avoid dangerous behaviors, we find significant effects on peer pressures from friends social influence status, i.e., given friends make the decision of not conducting dangerous behavior, a student benefits more from his conformity, or pays more for his disobedience,  if his friends are of high social influence than friends are low social influence.

We also compare results from our model with two other models: the constant peer effects (CPE) model in \cite{xu2011social} and the standard Logit  model. Note that the CPE model is nested in our model. Moreover, our model and the CPE model are structural approaches while the Logit model is of reduced--form. The coefficient estimates for age and gender are quite similar across three models. Effects from own social influence status are similar in our model and the Logit model. Moreover, both $\phi_1$ and $\psi_0$ are statistically significant at the 5\% 
level. In contrast, peer effects are insignificant in the CPE model. We also include the average friends' relative social influence into the Logit model, even an economics interpretation for its coefficient is implausible. The estimate of its coefficient is negative and statistically significant at the 5\% level. Such a result suggests  a negative correlation between players' decisions and their friends social influence status.

\begin{table}
\small
\caption{Estimation Results}
\label{table6}
{\begin{center}
\begin{tabular}{l|cccccc}
                        & Our model  & CPE model &Logit model\\\hline
Age                  &-0.020                                      &-0.006                &-0.018 \\
                        & (0.026)                                         &(0.025)                     &(0.026)   \\
Female            &  -0.778**                                       &-0.783**                    &-0.796**\\
                        &(0.084)                                          &(0.083)                      &(0.083)\\
Own $S_i$       &-0.404**                                         &-0.162**                    &-0.309**\\
                        &(0.095)                                          &(0.075)                     &(0.086) \\
Ave. Friends' $\bar S_{ji}$ &----                                                  &----                             &-0.510**\\
                        &----                                                  &----                             &(0.140)\\
Constant          &0.551                                       &0.258                        &0.718* \\
                        &(0.423)                                          & (0.416)                    &(0.413)\\
                  \\
$\phi_1$           & 1.694**                                         &   ----                           &  ----    \\
                         &(0.603)                                          &   ----                           &   ----   \\
$\psi_0$           & 0.707**                                          &0.369                  &    ---  \\
                        &(0.315)                                            &(0.310)                             &   ----     \\
$\psi_1$           &0.774                                        &  ----                            &     ----  \\
                        &(0.764)                                           & ----                             & ----  \\
$\phi_1+\psi_1$&2.468**                                           &  ----                            &   ----  \\
                         &   (1.335)                                                   &  ----                           &  ---- \\\hline
\multicolumn{3}{l}{\footnotesize{a. ** for 5\% significant; * for 10\% significant.}}\\
\multicolumn{3}{l}{\footnotesize{b. significance of peer effects obtains from the one--sided test.}}
\end{tabular}
\end{center}}

\end{table}

\bibliographystyle{econometrica}
\bibliography{overall}

\clearpage
\appendix
\small

\section{Proofs} 
\subsection{Proof of \Cref{thm1}}
\begin{proof}
\label{proofthm1}
We show  by contradiction. Let  $\Sigma^*(\mathbb W)=(\sigma^*_1(\mathbb W),\cdots,\sigma^*_n(\mathbb W))'$ and $\Sigma^\dag(\mathbb W)=(\sigma^\dag_1(\mathbb W),\cdots,\sigma^\dag_n(\mathbb W))'$ be two different profiles of equilibrium choice probabilities. Suppose  $\Sigma^*(\mathbb W)\neq\Sigma^\dag(\mathbb W)$. For $\Sigma\in [0,1]^n$, let
\[
\Gamma_i(\Sigma,\mathbb W)\equiv \frac{ \exp\left\{\beta(X_i)+\sum_{j\in F_i}\bar \alpha(S_j- S_i)\times \sigma_j(\mathbb W)\right\}}{ 1+\exp\left\{\beta(X_i)+\sum_{j\in F_i}\bar\alpha(S_j-S_i)\times \sigma_j(\mathbb W)\right\}}.
\]

By definition, $\Sigma^*(\mathbb W)$ and $\Sigma^\dag(\mathbb W)$ are two different solutions to the following equation:
\[
\sigma_i=\Gamma_i\left(\Sigma, \mathbb W\right), \ \ \forall \  i=1,\cdots,n.
\]
For player $i$, note that 
\begin{multline*}
\sigma_i^{*}(\mathbb W)-\sigma_i^{\dag}(\mathbb W)=\sum_{j\in F_i}\frac{\partial \Gamma_i(\tilde \Sigma, \mathbb W)}{\partial \sigma_j} \times \big[\sigma_j^{*}(\mathbb W)-\sigma_j^{\dag}(\mathbb W)\big]\\
=\sum_{j\in F_i}\Gamma_i(\tilde \Sigma, \mathbb W)\times [1-\Gamma_i(\tilde \Sigma, \mathbb W)]\times \bar\alpha(S_j-S_i)\times\big[\sigma_j^{*}(\mathbb W)-\sigma_j^{\dag}(\mathbb W)\big]
\end{multline*}
where $\tilde \Sigma$ is a choice probability between $\Sigma^{*}(\mathbb W)$ and $\Sigma^{\dag}(\mathbb W)$. Because $\Gamma_i\in(0,1)$, then $\Gamma_i\times (1-\Gamma_i)\leq 1/4$.  Therefore,
\begin{align*}
\left|\sigma_i^{*}(\mathbb W)-\sigma_i^{\dag}(\mathbb W)\right|&\leq \frac{1}{4}\sum_{j\in F_i} \left| \bar\alpha(S_j-S_i)\times \big[\sigma_j^{*}(\mathbb W)-\sigma_j^{\dag}(\mathbb W)\big]\right|\\
&\leq \frac{1}{4}\max_{j\in F_i}|\sigma_j^{^*}(\mathbb W)-\sigma_j^{\dag}(\mathbb W)|\times  \sup_{s\in \mathcal S_0}|\alpha(s)|
\end{align*}
By \Cref{ass2}, we have
\[
\left|\sigma_i^{*}(\mathbb W)-\sigma_i^{\dag}(\mathbb W)\right|<\max_{j\in F_i}\left| \sigma_j^{*}(\mathbb W)-\sigma_j^{\dag}(\mathbb W)\right|.
\]
Therefore,
\[
\max_{i\in\{1,\cdots,n\}}\left| \sigma_i^{*}(\mathbb W)-\sigma_i^{\dag}(\mathbb W)\right|<\max_{j\in \{1,\cdots,n\}}\left| \sigma_j^{*}(\mathbb W)-\sigma_j^{\dag}(\mathbb W)\right|
\]
leads to a contradiction. 
\end{proof}
\subsection{Proof of \Cref{thm2}}
\begin{proof}
\label{proofThm2}
Let 
\[
L_n(\theta, \Sigma)=\frac{1}{n}\sum_{i=1}^n\mathbb E \big\{Y_i\ln \Gamma_i (\Sigma, \theta;\mathbb W)+(1-Y_i)\ln \left[1-\Gamma_i(\Sigma,\theta; \mathbb W)\right]\big\}.
\] and $\Lambda_n=\{\theta\in\Theta:  \theta=\argmax_{c\in\Theta} L_n(c , \Sigma( \theta;\mathbb W))\}$.

Note that $L_n(\cdot,  \Sigma( \cdot;\mathbb W))$ is a continuously differentiable function in $\theta\in\Theta$. Because
\[
\mathbb E \big[Y_i\ln \Gamma_i (\Sigma, \theta;\mathbb W)\big]=\mathbb E \big[\sigma^*_i(\mathbb W)\ln \Gamma_i (\Sigma, \theta;\mathbb W)\big],
\]which is a continuously differentiable function of $\theta\in\Theta$ with bounded derivatives (uniformly over $n$), thus $L_n(\cdot,  \Sigma( \cdot;\mathbb W))$ uniformly converges to $L(\cdot,  \Sigma( \cdot;\mathbb W))$ under \Cref{ass8}.  It follows that  $\Lambda_n\rightarrow \{\theta_0\}$ as $n\rightarrow\infty$.

Moreover, following  \cite{xu2011social}, we have 
\[
\sup_{\theta\in \Theta} |\hat L_n(\theta, \Sigma)-L_n(\theta, \Sigma)|\overset{p}{\rightarrow} 0.
\] Further, by the argument in \cite{aguirregabiria2007sequential}, 
\[
d_{\mathcal H}(\hat \theta_{NPLE},\Lambda_n)\overset{p}{\rightarrow} 0,
\]where $d_{\mathcal H}$ is the Hausdorff--distance measure.  Therefore, $\hat \theta_{NPLE}\overset{p}{\rightarrow} \theta_0$.
\end{proof}

\subsection{Proof of \Cref{thm3}} 
\begin{proof}
\label{proofThm3}
From the first order condition we have that
\[
\frac{\partial \hat L_n(\hat\theta_{NPLE},\Sigma(\hat\theta_{NPLE};\mathbb W))}{\partial \theta}=0.
\]Take Taylor expansion on the above equation around the true parameter $\theta_0$, we have
\begin{multline*}
\frac{\partial \hat L_n(\theta_0,\Sigma(\theta_0;\mathbb W))}{\partial \theta}+\left[\frac{\partial^2 \hat L_n(\theta_0,\Sigma(\theta_0;\mathbb W))}{\partial \theta\partial \theta'}
+\frac{\partial^2 \hat L_n(\theta_0,\Sigma(\theta_0;\mathbb W))}{\partial \theta\partial \Sigma}\frac{\partial \Sigma(\theta_0;\mathbb W)}{\partial \theta}\right](\hat\theta_{NPLE}-\theta_0)\\
=O_p(n^{-1}).
\end{multline*}
Note that
\[
\frac{\partial \hat L_n(\theta,\Sigma)}{\partial \theta}=\frac{1}{n}\sum_{i=1}^n Z_i(Y_i-\Gamma_i(\Sigma,\theta;\mathbb W))
\]Therefore,
\[
\frac{\partial^2 \hat L_n(\theta_0,\Sigma)}{\partial \theta\partial \theta'}=-\frac{1}{n}\sum_{i=1}^n Z_iZ_i'\sigma_i(1-\sigma_i)
\]and
\[
\frac{\partial^2 \hat L_n(\theta,\Sigma(\theta;\mathbb W))}{\partial \theta\partial \Sigma}\frac{\partial \Sigma(\theta;\mathbb W)}{\partial \theta}=-\frac{1}{n}\sum_{i=1}^n \sum_{j\in F_i}Z_i Z_j'\sigma^*_j(1-\sigma^*_j) \times \frac{\psi_0+(\phi_1+\psi_1) \bar S_{ji}}{Q_i}.
\]
Therefore,
\begin{multline*}
\left[\frac{\partial^2 \hat L_n(\theta_0,\Sigma(\theta_0;\mathbb W))}{\partial \theta\partial \theta'}
+\frac{\partial^2 \hat L_n(\theta_0,\Sigma(\theta_0;\mathbb W))}{\partial \theta\partial \Sigma}\frac{\partial \Sigma(\theta_0;\mathbb W)}{\partial \theta}\right]\times \sqrt n(\hat\theta_{NPLE}-\theta_0)\\
=-\frac{1}{\sqrt n}\sum_{i=1}^nZ_i'\{Y_i-\sigma^*_i(\mathbb W)\}+o_p(1)
\end{multline*}
Because $Y_i$ is conditionally independent (conditional on $W_n$), by conditional central limit theorem  \citep[see e.g.][]{van2000asymptotic} and \Cref{ass10}, we have
\[
\sqrt n(\hat\theta_{NPLE}-\theta_0)\xrightarrow{d}N(0,\Omega(\theta_0))
\]where $\Omega(\theta_0)$ is given by \Cref{thm3}
\end{proof}
\end{document}